\documentclass[letterpaper]{article} 
\usepackage{aaai23}  
\usepackage{times}  
\usepackage{helvet}  
\usepackage{courier}  
\usepackage[hyphens]{url}  
\usepackage{graphicx} 
\usepackage{natbib}  
\usepackage{caption} 
\usepackage{algorithm}
\usepackage{algorithmic}
\usepackage{subcaption}

\usepackage{newfloat}
\usepackage{listings}
\DeclareCaptionStyle{ruled}{labelfont=normalfont,labelsep=colon,strut=off} 
\floatstyle{ruled}
\newfloat{listing}{tb}{lst}{}
\floatname{listing}{Listing}
\usepackage{eucal}
\usepackage{xcolor}
\usepackage{kotex}
\usepackage{multirow}
\usepackage{xcolor}
\usepackage{graphicx}
\usepackage{comment}

\title{Music Playlist Title Generation Using Artist Information}
\author {
    Haven Kim,\textsuperscript{\rm 1}
    Seungheon Doh, \textsuperscript{\rm 1}
    Junwon Lee, \textsuperscript{\rm 2}
    Juhan Nam \textsuperscript{\rm 1}
}
\affiliations {
    \textsuperscript{\rm 1}Graduate School of Culture Technology, KAIST\\
    \textsuperscript{\rm 2}Department of Electrical Engineering, KAIST\\
    khaven@kaist.ac.kr,
    seungheondoh@kaist.ac.kr,
    james39@kaist.ac.kr,
    juhan.nam@kaist.ac.kr
}

\usepackage{mathrsfs}
\usepackage{bibentry}
\usepackage{amsmath}
\usepackage{booktabs}

\begin{document}

\maketitle

\begin{abstract}
Automatically generating or captioning music playlist titles given a set of tracks is of significant interest in music streaming services as customized playlists are widely used in personalized music recommendation, and well-composed text titles attract users and help their music discovery. We present an encoder-decoder model that generates a playlist title from a sequence of music tracks. While previous work takes track IDs as tokenized input for playlist title generation, we use artist IDs corresponding to the tracks to mitigate the issue from the long-tail distribution of tracks included in the playlist dataset. Also, we introduce a chronological data split method to deal with newly-released tracks in real-world scenarios. Comparing the track IDs and artist IDs as input sequences, we show that the artist-based approach significantly enhances the performance in terms of word overlap, semantic relevance, and diversity.

\end{abstract}

\section{Introduction}
With the popularization of music streaming services, music playlists have become essential resources in helping music discovery from huge collections of music \cite{dias2017}. Because of the expected difficulty of browsing a large number of music playlists, a music playlist title that depicts the content information is often provided to attract users and help their discovery. For example, a title could be used to describe genre (e.g., ``\textit{send this playlist to your friend who doesn't know what indie rock is}''), situation (e.g., ``\textit{best jogging music for motivation}''), originality (e.g., ``\textit{playlist that will make you feel like an Arabian princess}''), and so on. Making an informative and attractive title, however, does not come easily to everybody; we have seen a lot of titles that do not contain musically meaningful information (e.g., ``\textit{my favorites}'') or simply consist of a few related words (e.g., ``\textit{meditation, zen, yoga, sleep, study, focus}''). Thus, if a descriptive playlist title could be generated automatically, it would help make the playlist more inviting and discoverable.


Several approaches have been proposed to automatically generate a playlist's title or description. They are mainly based on a neural machine translation model, which encodes a whole sequence of musical information into fixed-length latent vectors and then decodes them into text \cite{bahdanau2014neural,vaswani2017}. 
Specifically, they used an RNN-based sequence-to-sequence model that learns from audio and supplementary text data (e.g., genre) \cite{choi2016} or a transformer-based model that takes a series of track IDs (unique IDs given to each track) \cite{doh2021} to generate a title for a music playlist. 
From the previous work, we observe two major issues. One is that a significant amount of the data is rarely-seen, making it difficult to capture the semantics for the model. As a result, the model has a tendency to repetitively produce musically meaningless titles such as ``\textit{my favorite songs}''. Another issue is that it does not reflect real-world settings in which the model makes inferences from playlists containing new releases. An ideal music playlist title generation system should generalize well across different types of music playlists; its performance should not deviate when inferring from playlists with rarely appearing tracks or new releases.

To address these limitations, we introduce a music playlist title generation model that takes artist IDs instead of track IDs as input (Section 2). Because tracks by the same artist all use the same input embedding vector, the model is capable of capturing the semantics of songs regardless of their frequency of appearance in playlists. To better represent real-world settings, we propose a chronological data split method to simulate the case of inferring from tracks that are newly released (Section 3). Furthermore, we perform a systematic evaluation regarding word overlap, semantic relevance, and diversity (Section 4). Section 5 shows that the artist ID embedding approach improves the ability to deal with unpopular tracks or artists without compromising the ability to handle popular tracks or artists. Finally, Section 6 concludes this paper by discussing its limitations and future work. Reproducible code is available online \footnote[1]{ \texttt{github.com/havenpersona/title\_generation}}.

\begin{figure}[!t]
   \centering
   \includegraphics[width=1\linewidth]{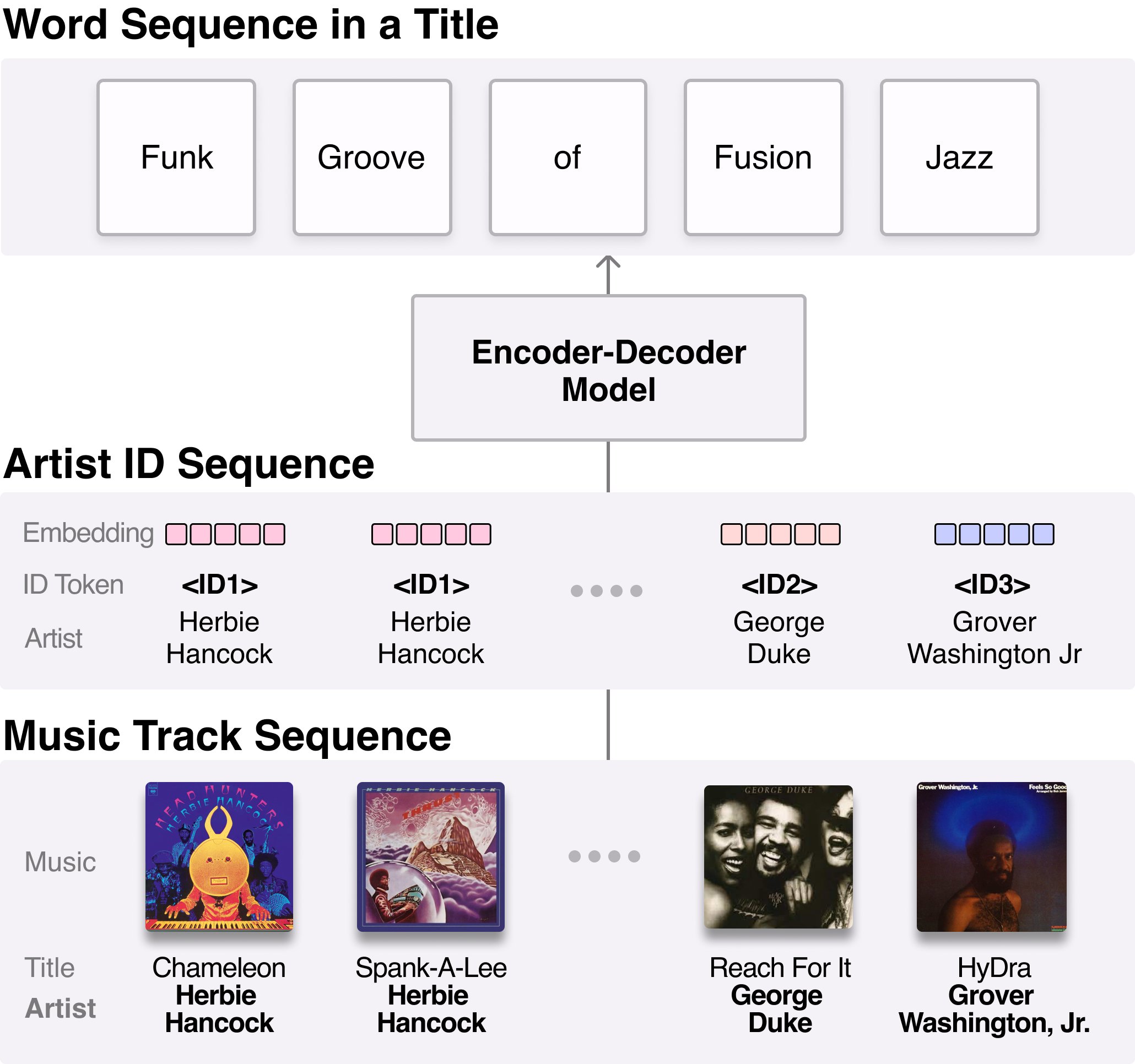}
   \caption{Model Architecture.}
   \label{fig:teaser}
\end{figure}

\section{Music Playlist Title Generation}
\subsection{Encoder-Decoder Model}

Similar to the previous work \cite{doh2021}, our music playlist title generation model is based on an encoder-decoder framework: an encoder network for obtaining the semantics of music tracks and a decoder for generating a sequence of natural language. We use the Transformer architecture \cite{vaswani2017}, which has two or three sub-layers: multi-head attention sub-layers, fully-connected sub-layers, and the masked multi-head attention layer in the case of the decoder. As illustrated Figure~\ref{fig:teaser}, the architecture takes unique token IDs from a sequence of music tracks. In the proposed systems, a music track sequence is a collection of track or artist information $x$, with each track or artist represented as a unique token $(x_1, ..., x_n)$. The tokenized IDs are initialized by $d_{model}$~dimension random vectors and processed by the encoder network $f\colon \mathcal{X}\mapsto \mathcal{Z}$. The vectors are transformed into a sequence of latent representations $z = (z_1, ..., z_n)$, which is fed to the decoder $g\colon$ $\mathcal{Z} \mapsto \mathcal{Y}$, and it generates an output sequence $\hat{y} = (\hat{y}_{1},..,\hat{y}_{m})$ where $\hat{y}$ refers to a generated music playlist title and each element in $(\hat{y}_{1},..,\hat{y}_{m})$ refers to each token in the title. We used the softmax cross-entropy loss for comparing the ground truth with the prediction.

 
\subsection{Music Track Representations}
We use music track sequences as input for the encoders. Each element of the track sequence was represented with metadata, such as a track ID and an artist ID. Tracks provide more specific information than artists. Such high specificity may capture the detail of music semantics, but it often causes the issue that the system cannot extract meaningful semantics if the track was infrequently seen in the training phase. For example, in the playlist datasets in our experiment, more than 17\% of tracks do not appear more than two times in the training phase.
On the other hand, only less than 5.0\% artist IDs appear more than twice.
Therefore, using artist IDs may allow the system to extract the semantics from the information obtained from prior songs by the same artists. 

\begin{table}[!t]
\centering
\resizebox{\linewidth}{!}{%
\begin{tabular}{l|l|c|c|c}
\toprule
\textbf{Dataset} & \textbf{Statistic} & \textbf{Original} & \textbf{Doh et al.} & \textbf{Proposed} \\ \midrule
\multirow{8}{*}{\begin{tabular}[c]{@{}l@{}}Melon \end{tabular}} & \# of Playlist & 148,826 & 51,404 & 80,160 \\ 
 & \# of Unique Track & 649,902 & 429,266 & 516,220 \\ 
 & \# of Unique Artist & 115,457 & 77,145 & 92,020 \\ 
 & \# of Unique Title & 115,318 & 59,209 & 77,905 \\ 
 & \# of Unique Word & 88,524 & 49,978 & 62,834 \\ 
 & Avg. Char Length & 2.8 & 3.6 & 3.1 \\ 
 & Avg. Title Length & 3.6 & 4.7 & 5.2 \\ 
 & Avg. Track Length & 39.7 & 46.2 & 44.6 \\ \midrule
\multirow{8}{*}{\begin{tabular}[c]{@{}l@{}}MPD \end{tabular}} & \# of Playlist & 1,000,000 & 50,083 & 34,115 \\ 
 & \# of Unique Track & 2,262,292 & 402,523 & 303,779 \\ 
 & \# of Unique Artist & 286,787 & 69,641 & 55,103 \\ 
 & \# of Unique Title & 17,381 & 1,859 & 1,229 \\ 
 & \# of Unique Word & 11,146 & 1,886 & 1,258 \\ 
 & Avg. Char Length & 5.2 & 4.2 & 4.1 \\ 
 & Avg. Title Length & 1.4 & 3.4 & 3.4 \\ 
 & Avg. Track Length & 66.3 & 66.3 & 64.2 \\ \bottomrule
\end{tabular}
}
\caption{\label{tab:statistics}Dataset Statistics.}
\end{table}

\section{Dataset and Experiment}

We used two different datasets: the Melon Playlist Dataset (Melon) \cite{ferraro2021} and the Spotify Million Playlist Dataset (MPD) \cite{chen2018}. The dominant languages used are Korean and English, respectively. Originally created for the automatic playlist continuation (APC) tasks, the datasets contain playlist titles, music track sequences, and music metadata (track, artist name), along with the last modification date. Using both datasets, the previous work \cite{doh2021} suggested a random data split scheme and a noise-cleaning strategy for generating playlist titles. However, this approach has two limitations – First, validating and testing on randomly divided data does not mimic the circumstance of making inferences from newly released tracks and artists. Second, the filtering strategy considered the structural quality of titles only, while overlooking the semantic quality. This would reduce the usability of the system in practice. To address the problem, we propose a new data split and noise-filtering strategy.

\begin{table}[!t]
\centering
\begin{tabular}{c|c|c}
\toprule
\textbf{Dataset} & \textbf{Entity} & \textbf{\begin{tabular}[c]{@{}c@{}}\texttt{\textless{}UNK\textgreater} \\ Proportion \end{tabular}} \\ \midrule
\multirow{2}{*}{\begin{tabular}[c]{@{}c@{}} Melon \end{tabular}} & Track ID & 14.70\% \\ 
 & Artist ID & 3.53\% \\ \midrule
\multirow{2}{*}{\begin{tabular}[c]{@{}c@{}} MPD \end{tabular}} & Track ID & 9.86\% \\ 
 & Artist ID & 2.91\% \\ \bottomrule
\end{tabular}
\caption{\label{tab:unk}Proportion of \texttt{\textless{}UNK\textgreater} tokens in validation and test sets}
\end{table}

\begin{figure}[!t]
     \centering
    \begin{subfigure}[t]{0.23\textwidth}
        \raisebox{-\height}{\includegraphics[width=\textwidth]{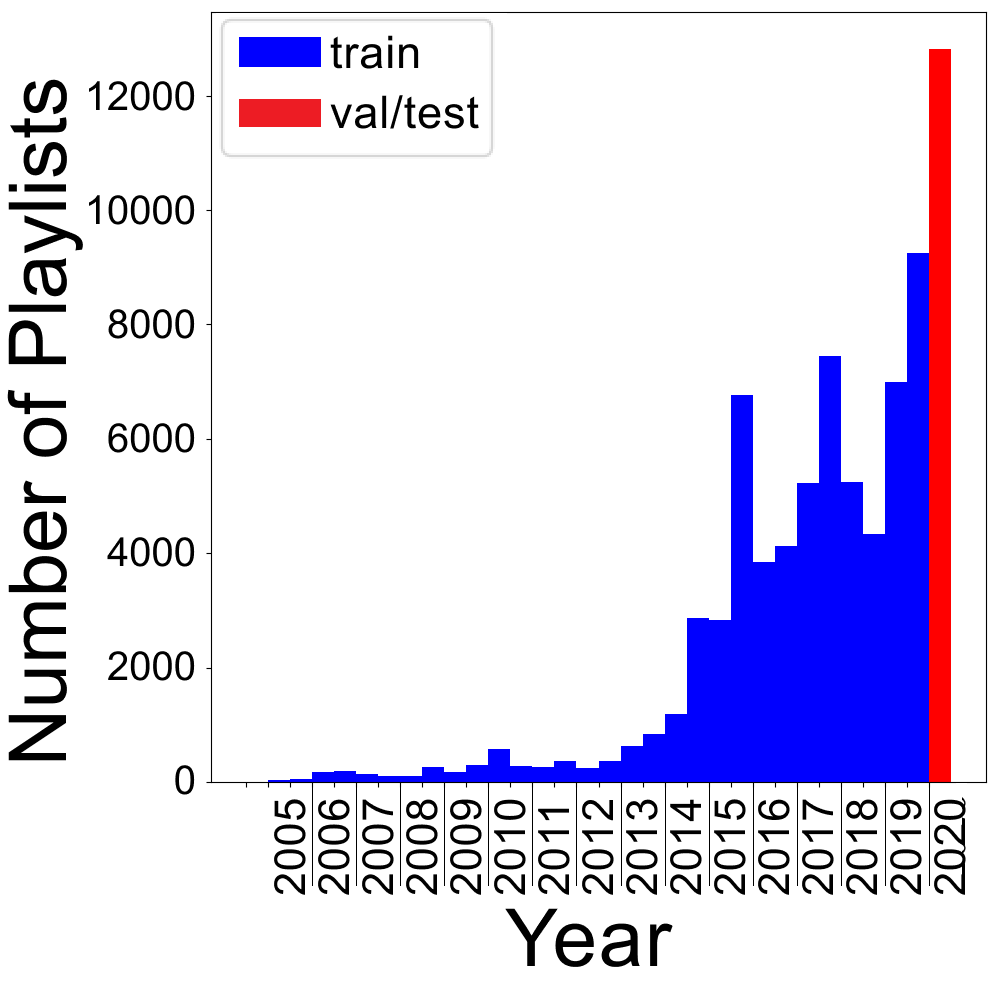}}
        \caption{Melon Playlist}
    \end{subfigure}
    \begin{subfigure}[t]{0.23\textwidth}
        \raisebox{-\height}{\includegraphics[width=\textwidth]{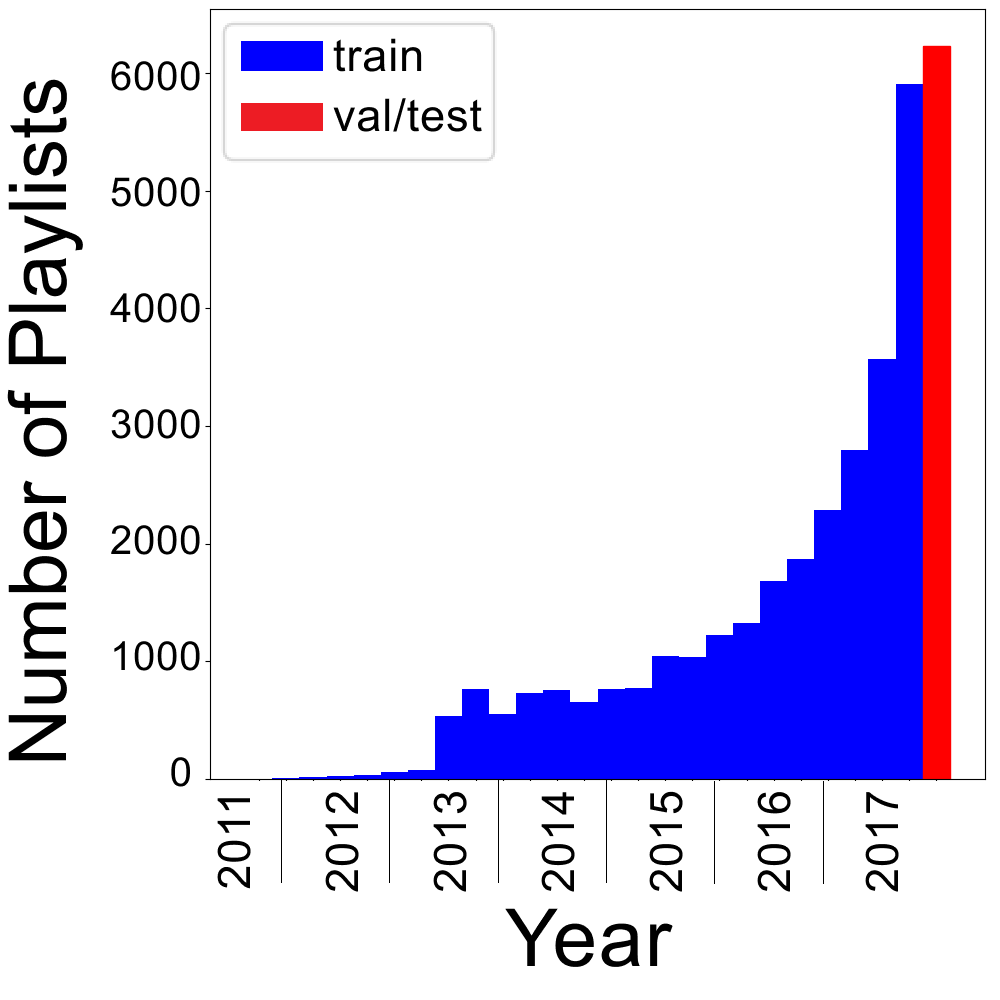}}
        \caption{Million Playlist}
    \end{subfigure}
    \caption{Chronologically split data. Training data is colored blue, and validation/test data are colored red. The horizontal axes are split by year and quarter.}
    \label{fig:chronology}
\end{figure}


\subsection{Chronological Data Split}
In the real world, the model makes predictions on playlists with new releases as a large number of music tracks are uploaded to music streaming platforms every day. Validating and testing on randomly split sets does not represent this reality. To overcome this limitation, we performed training, validation, and testing on data that was chronologically divided, as shown in Figure \ref{fig:chronology} and replaced a new release with a special token, \texttt{\textless{}UNK\textgreater}. For the Melon Playlist Dataset, we used playlists made in 2020 as validation and test sets. Any playlist created before that was used as training data. For the Million Playlist Dataset, playlists made in the last quarter of 2017 were used as validation and test sets. Those created before that were used as training data. The approximated ratio of train sets to validation/test sets is 83:17 and 82:18, respectively.  As Table \ref{tab:unk} shows, the proportion of \texttt{\textless{}UNK\textgreater} token is significantly higher when using track IDs as input than artist IDs, which implies that the model is more likely to handle unseen data when taking track IDs as input.

\begin{table*}[!t]
\centering
\begin{tabular}{l|l|c|ccccc|cc}
\toprule
\multirow{2}{*}{\textbf{Dataset}} & \multirow{2}{*}{\textbf{Input}} & \textbf{Loss} & \multicolumn{5}{c|}{\textbf{n-gram Overlap Metrics}} & \multicolumn{2}{c}{\textbf{BERT-Based Metrics}} \\
 &  & NLL & BLEU-1 & BLEU-2 & ROUGE-1 & ROUGE-2 & METEOR & BERT Score & SentBERT \\ \midrule
\multirow{2}{*}{Melon} & Track ID & 6.40 & 0.048 & 0.01 & 0.061 & 0.01 & 0.041 & 0.417 & 0.775 \\
 & Artist ID & 6.23 & 0.056 & 0.01 & 0.074 & 0.01 & 0.049 & 0.430 & 0.778 \\  \midrule
\multirow{2}{*}{MPD} & Track ID & 1.35 & 0.232 & 0.180 & 0.237 & 0.174 & 0.212 & 0.889 & 0.408 \\
 & Artist ID & 1.17 & 0.288 & 0.229 & 0.293 & 0.223 & 0.265 & 0.897 & 0.460 \\ \bottomrule
\end{tabular}
\caption{Comparison between the track ID embedding model and the artist ID embedding model. The results indicate that word overlap and semantic relevance improve when using the artist ID embedding.}
\vspace{3mm}
\label{tab:embeddings}
\end{table*}

\begin{figure*}
     \centering
    \begin{subfigure}[t]{0.49\textwidth}
        \raisebox{-\height}{\includegraphics[width=\textwidth]{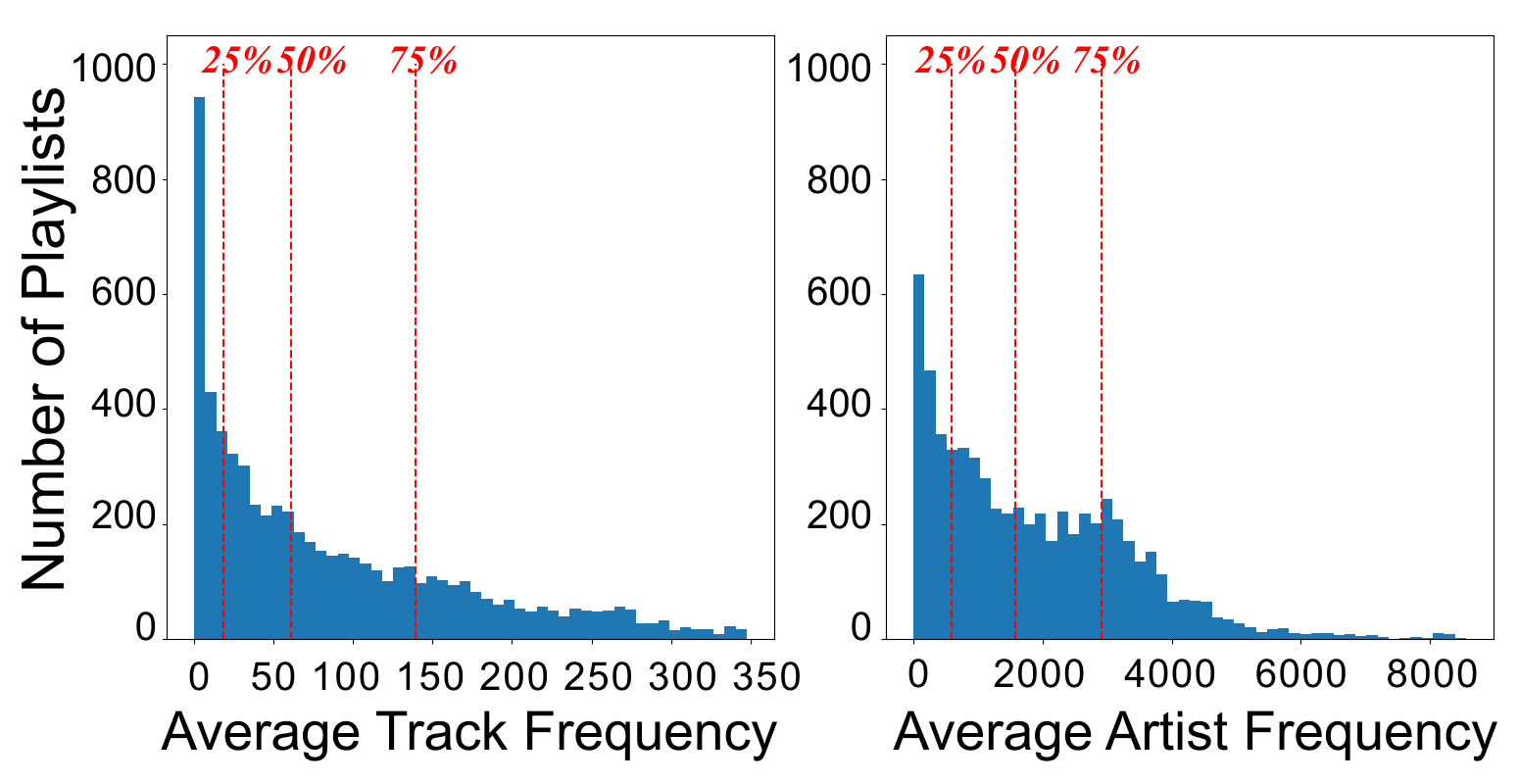}}
        \caption{Melon Playlist}
    \end{subfigure}
    \begin{subfigure}[t]{0.49\textwidth}
        \raisebox{-\height}{\includegraphics[width=\textwidth]{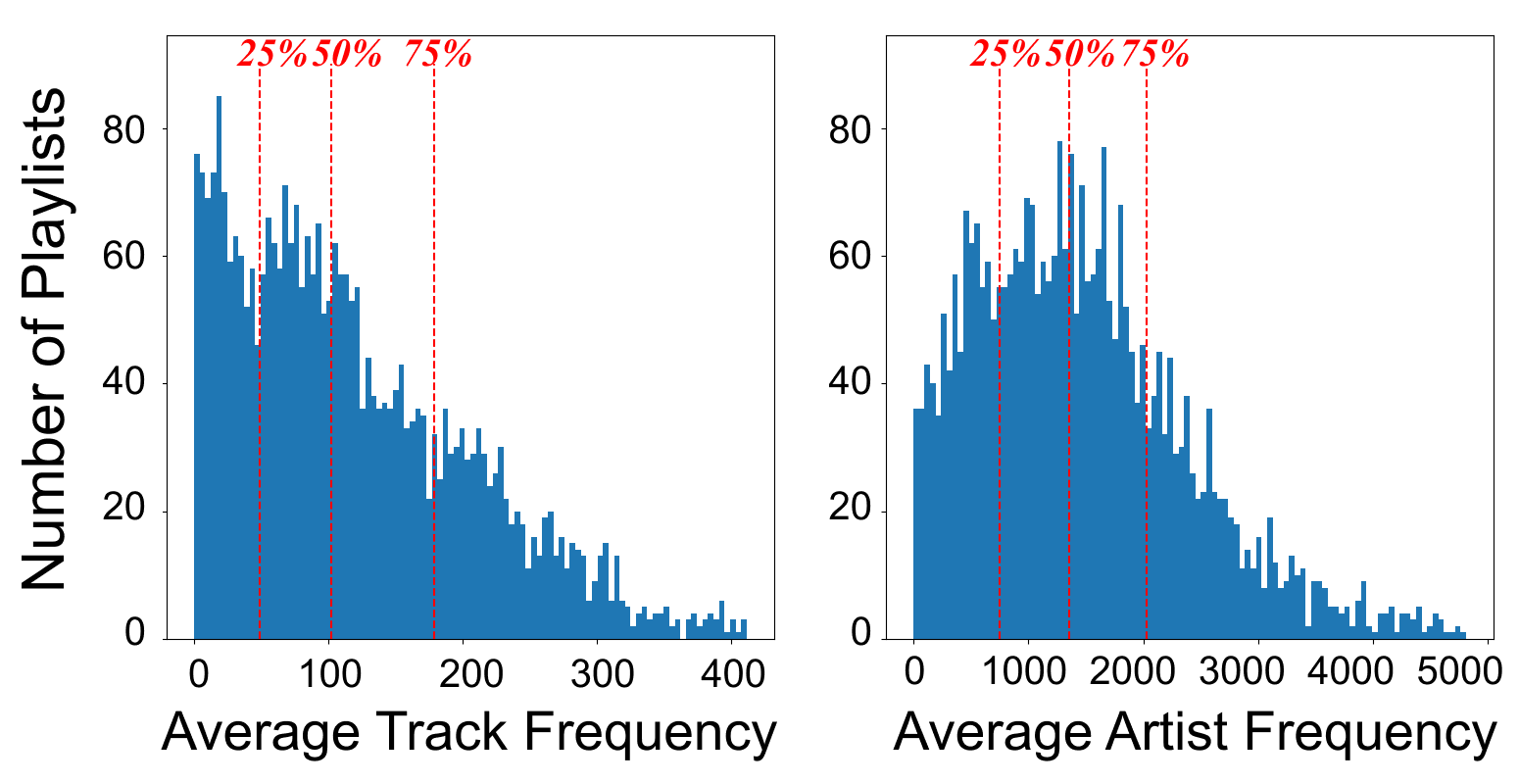}}
        \caption{Million Playlist}
    \end{subfigure}
    \caption{Distribution of average track frequency ($F_t$) and average artist frequency ($F_a$). We removed items above the 99th percentile to eliminate statistical outliers. The red lines were added to visualize the values that divide the data into four parts.}
    \label{fig:popularity_distribution}
\vspace{2mm}
\end{figure*}

\subsection{Filtering Out Noisy Data}

As the previous study \cite{doh2021} suggested, some playlists in datasets come with undesirable titles or a very small number of tracks (including playlists with no tracks). Therefore, they only used playlists that met the following criteria: i) The title should have at least three tokens after white space tokenization. ii) The average character length of tokens should be at least three. iii) The playlist should contain at least ten tracks. We modified the criteria by mitigating the second and third criteria and adding a newly introduced one. The new criterion was introduced to avoid generating titles without musically meaningful information, such as \textit{``good music to listen to''} or \textit{``best music ever''}. Therefore, we only used playlists that contained at least one word from a pre-defined list of music tags. The following steps were taken to obtain the list of music tags.

\textbf{Korean}: The list of music tags in Korean is based on the 30,652 unique tags provided by the Melon Playlist Dataset. We only considered tags that appeared at least fifty times in the entire dataset. Tags that are musically meaningless, such as \textit{``recommendation''} and \textit{``compilation''} were excluded. The total number of tags in Korean is 731.

\textbf{English}: The Million Playlist Dataset does not provide music tags. We obtained the tag list in English from two publicly available music tag datasets for music auto-tagging research (the MTG-Jamendo Dataset \cite{jamendo} and the Million Song Dataset \cite{msd}). Since they do not contain a large number of genres that have recently gained popularity (e.g., \textit{``trap''}), we gained additional tags from frequently appearing words on YouTube Music. Again, we removed tags that are not suitable for music playlist descriptions. As a result, we obtained a total of 1,013 English tags.



To summarize, we filtered out playlists that did not meet the following criteria: i) The title should have at least three tokens after white space tokenization. ii) The average character length of tokens should be at least two. iii) The playlists should contain at least two tracks. iv) The title should contain at least one musically meaningful word. As a result of our modification of criteria, we were able to collect titles with meaningful information, as all of the titles include at least one musically relevant term without reducing the average title length (see Table \ref{tab:statistics}).

\subsection{Training Details}
The model was optimized with the Adam optimizer with a 0.005 learning rate and $10^{-4}$ learning rate decay \cite{adam_optimizer}. The rate was scheduled with the cosine annealing scheduler with a minimum learning rate of $10^{-6}$ \cite{cosine2017}. We used a batch size of 64 and performed validation at the end of every epoch. Following previous research, we ignored the position of the input sequence because the order of tracks or artists in a music playlist does not provide crucial information when generating a title in most circumstances.

\section{Evaluation Metrics} 

For our evaluation, we use text generation evaluation metrics \cite{celikyilmaz2020evaluation} and grouped them into three categories: (i) $n$-gram based overlap metrics (ii) BERT-based neural metrics (iii) $n$-gram based diversity metrics. Using these metrics, we assess the overall performance. In addition, we measure the robustness which we define as the degree to which the model's performance changes when dealing with data that appears infrequently versus frequently.

\subsection{$n$-gram Overlap Metrics}
Counting the number of overlapping $n$-gram between a reference $y = (y_1, ..., y_l)$ and a candidate $\hat{y} = (\hat{y}_1, ..., \hat{y}_m)$ is one of the most common ways to evaluate models for text generation. This paper uses three popular $n$-gram overlap metrics to evaluate the model performance — BLEU, ROUGE, and METEOR.

\textbf{Bilingual Evaluation Understudy (BLEU)} was proposed to assess machine translation performance, and this includes modified $n$-gram precision \cite{papineni}. The computation of the modified precision score for one reference sentence and one candidate sentence is done by counting the number of $n$-gram matches and dividing it by the number of candidate $n$-grams. In this paper, we calculated BLEU for two different values of $n$ ($n = 1, 2$).

\textbf{Recall-Oriented Understudy for Gisting Evaluation (ROUGE)} was proposed for text summarization evaluation and has also been used for evaluating short text generation \cite{lin2004}. Whereas BLEU focuses on precision, ROUGE uses both recall and precision. In this paper, we use the f1-score of ROUGE-$N$, one of the variants of ROUGE, for evaluating our model. It is obtained by calculating the harmonic mean of prediction (the number of $n$-gram matches divided by the number of candidate $n$-grams) and recall (the number of $n$-gram matches divided by the number of reference $n$-grams). The values of $n$ used are 1 and 2.

\textbf{Metric for Evaluation of Translation with Explicit ORdering (METEOR)} was proposed to address some limitations of BLEU for machine translation evaluation \cite{banerjee2005}. In a similar fashion to ROUGE, this metric considers both recall and precision. It could be obtained by calculating the harmonic mean of the unigram precision score and the unigram recall score, with weighting recall nine times more than precision. The final score combines this with the chunk penalty. The penalty increases if the number of chunks (consecutive unigrams) increases. When there is no bi-gram match, the maximum number of penalty (0.5) is given.

\subsection{BERT-Based Neural Metrics}
Instead of counting the number of exact matches, BERT-based metrics use the pre-trained contextual embeddings generated by the model to capture the semantic similarity between a reference sentence $y = (y_1, ..., y_l)$ and a candidate sentence $\hat{y} = (\hat{y}_1, ..., \hat{y}_m)$.

\textbf{BERT Score} is obtained by computing the cosine similarity between each token in a reference sentence and in a candidate sentence and using greedy matching to maximize the score \cite{zhang2019bertscore}. In this paper, we used the pretrained \texttt{KcBert-Base} model \cite{lee2020kcbert} for evaluating the Melon Playlist Dataset and the pretrained \texttt{RoBERTa-large} model \cite{roberta2021} for evaluating the Million Playlist Dataset to get the contextual embeddings.

\textbf{Sentence BERT (SentBERT)} was proposed for semantic textual similarity evaluation tasks \cite{reimer_2019}. The SentBERT uses a siamese pre-trained BERT network to derive semantically meaningful sentence embeddings. Each embedding is calculated by taking the mean of all vector representations of each token and computing the cosine similarity between them. In this paper, we adopted the embeddings of \texttt{KLEU RoberTa small} \cite{klue2021} for evaluating the Melon Playlist Dataset and the \texttt{all-MiniLM-L6-v2} \cite{minilm2020} for evaluating the Million Playlist Dataset.

\begin{figure*}
     \centering
    \begin{subfigure}[t]{0.49\textwidth}
        \raisebox{-\height}{\includegraphics[width=\textwidth]{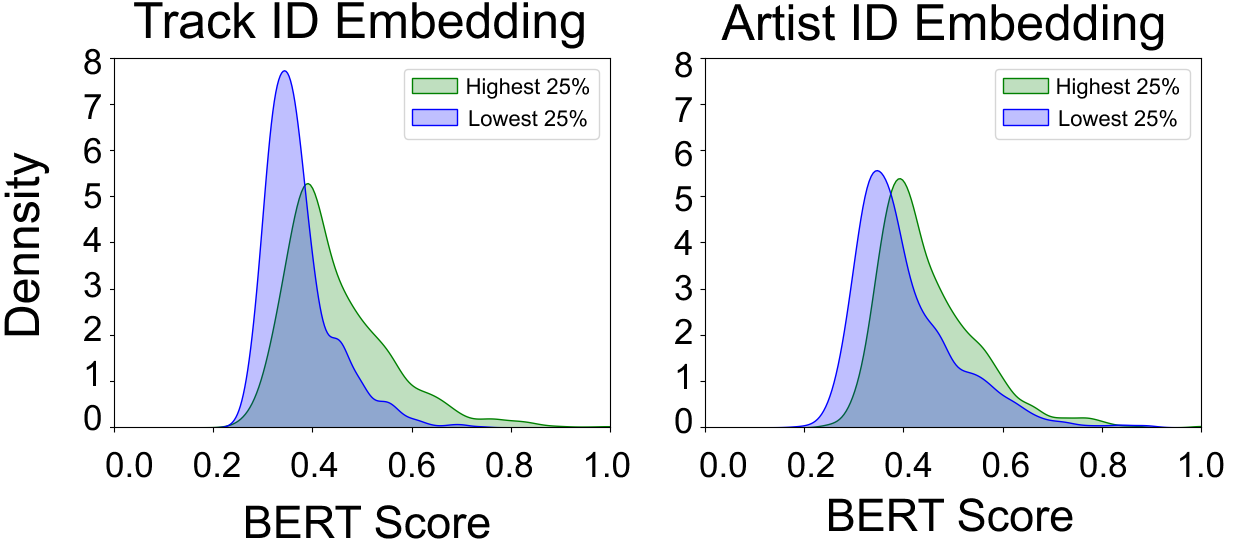}}
        \caption{Melon Playlist}
    \end{subfigure}
    \begin{subfigure}[t]{0.49\textwidth}
        \raisebox{-\height}{\includegraphics[width=\textwidth]{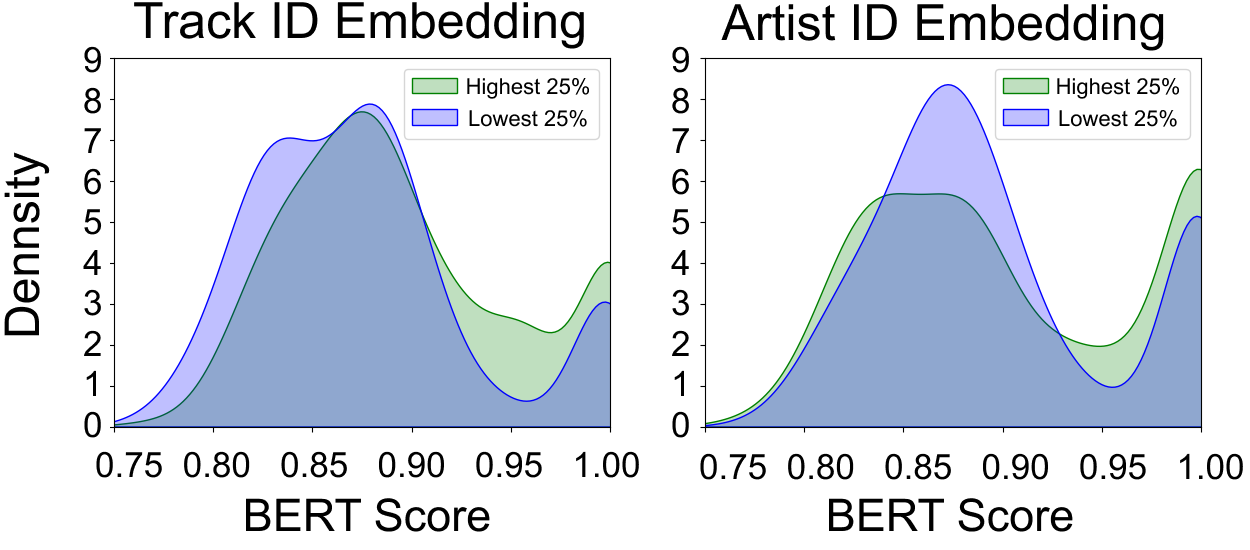}}
        \caption{Million Playlist}
    \end{subfigure}
    \caption{Distribution of BERT Score of the highest 25\% $F_t$ group and the lowest 25\% $F_t$ group. The score of the lowest 25\% group is improved without impairng the performance of the highest 25\% group when the artist ID embedding is used.}
    \label{fig:bert_score_f_t}
\end{figure*}

\begin{figure*}
     \centering
    \begin{subfigure}[t]{0.49\textwidth}
        \raisebox{-\height}{\includegraphics[width=\textwidth]{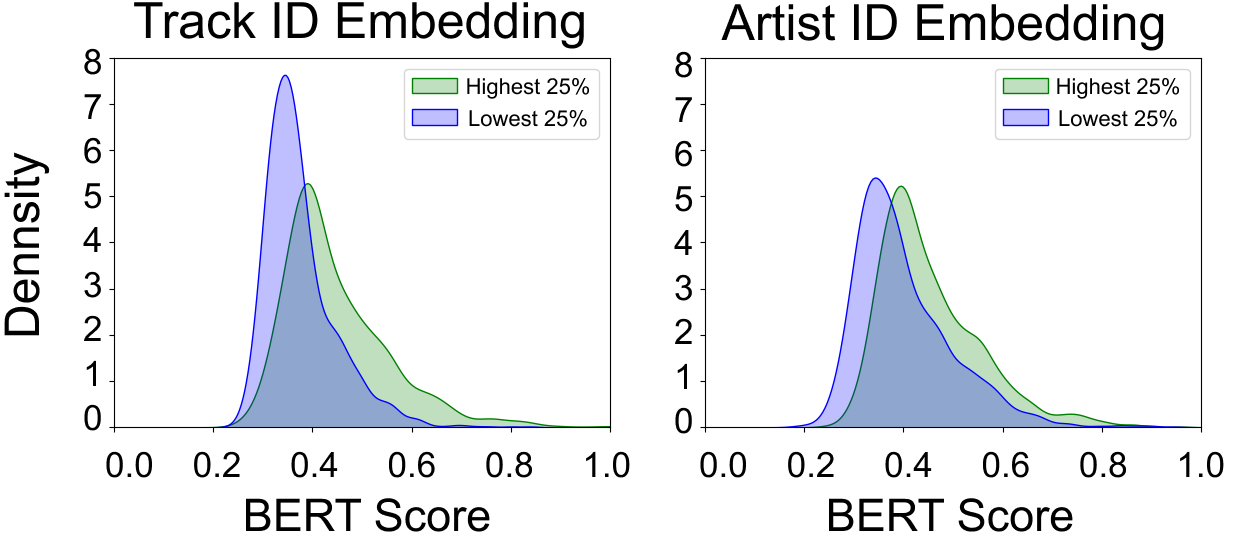}}
        \caption{Melon Playlist}
    \end{subfigure}
    \begin{subfigure}[t]{0.49\textwidth}
        \raisebox{-\height}{\includegraphics[width=\textwidth]{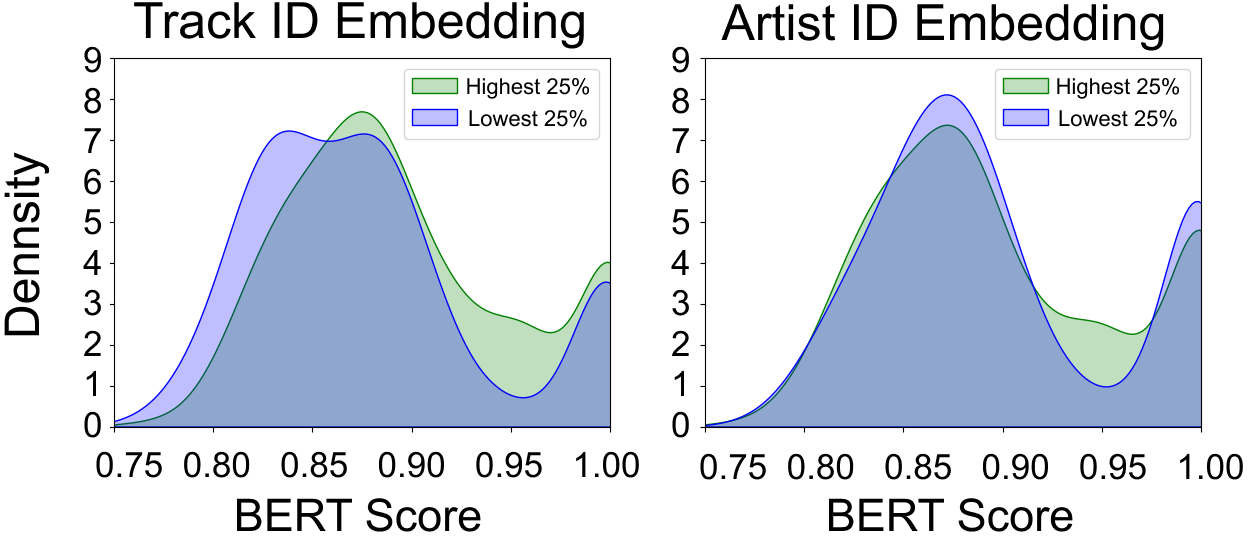}}
        \caption{Million Playlist}
    \end{subfigure}
    \caption{Distribution of BERT Score of the highest 25\% $F_a$ group and the lowest 25\% $F_a$ group. The score of the lowest 25\% group is improved without imparing the performance of the highest 25\% group when the artist ID embedding is used.}
    \label{fig:bert_score_f_a}
\end{figure*}

\subsection{$n$-gram-Based Diversity Metrics}
A successful music playlist title generation model should avoid generating repetitive titles and instead produce diverse outputs. Researchers suggested $n$-gram-based diversity metrics show a high correlation with lexical diversity \cite{tevet2021evaluating}. One of the $n$-gram based metrics used for diversity evaluation is \textit{distinct-$n$}, the number of distinct $n$-grams divided by the total number of $n$ grams \cite{diversity}. In this paper, we evaluated lexical variations using \textit{distinct-$n$} with three values of $n$ ($n = 1,2,3$).

\section{Results and Discussion}
\subsection{Word Overlap and Semantic Relevance}

Table \ref{tab:embeddings} compares all $n$-gram overlap performances (BLEU, ROUGE, and METEOR) of different input models (Track ID, Artist ID). The artist ID embedding model shows a clear performance improvement over the track ID embedding model. In the case of the Melon Playlist Dataset, the change in BLEU-2 was not noticeable because of the grammatical characteristics of Korean (that post-positional particles immediately follow a noun or pronoun without white space). We observed a similar trend on the Negative Log-likelihood (NLL) loss and the BERT-based neural metrics (BERT Score and Sentence BERT). This indicates that utilizing artist IDs lessens the problem of dealing with infrequently appearing tracks, enabling the system to extract more musically meaningful information.


\begin{figure}
     \centering
    \begin{subfigure}[t]{0.23\textwidth}
        \raisebox{-\height}{\includegraphics[width=\textwidth]{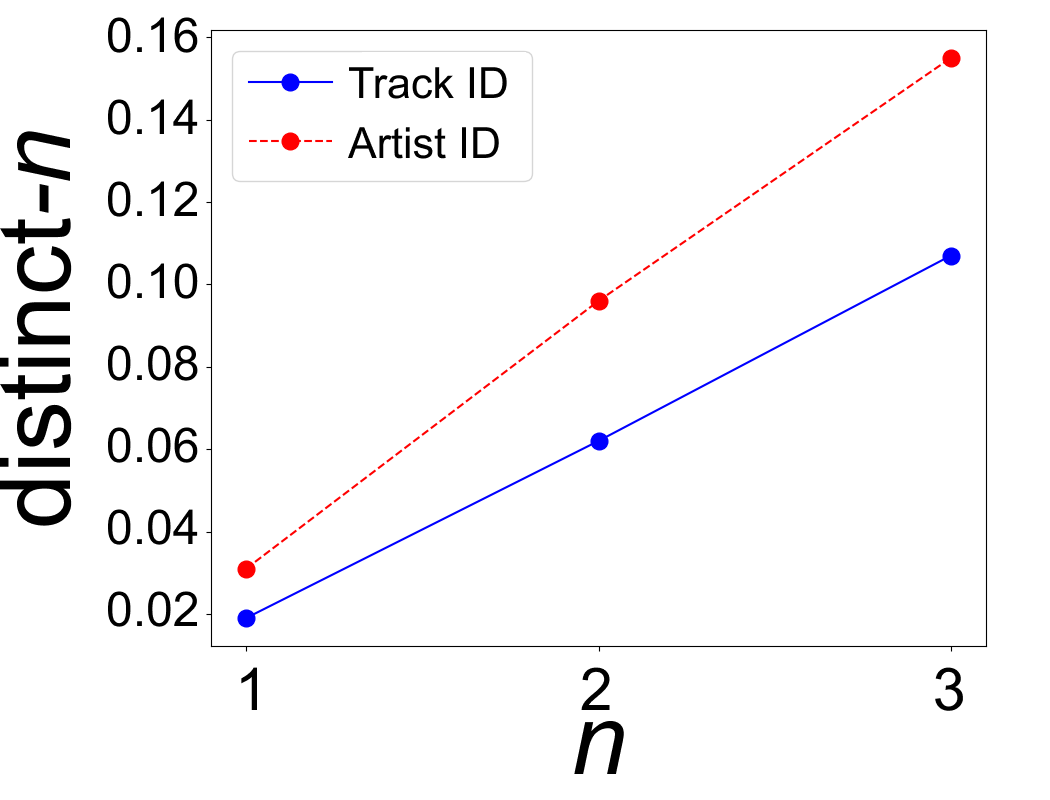}}
        \caption{Melon Playlist}
    \end{subfigure}
    \begin{subfigure}[t]{0.23\textwidth}
        \raisebox{-\height}{\includegraphics[width=\textwidth]{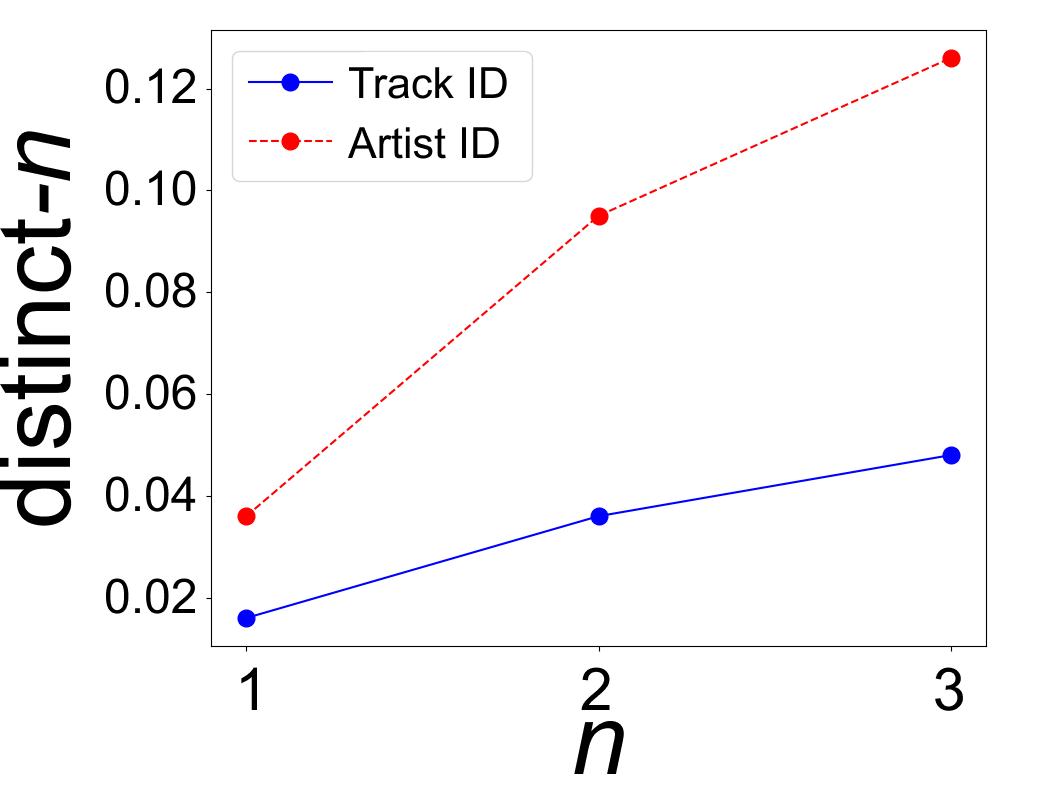}}
        \caption{Million Playlist}
    \end{subfigure}
    \caption{\textit{distinct-$n$} ($n = 1,2,3$) indicates that the artist ID embedding generates more lexically diverse titles.}
    \label{fig:diversity}
\end{figure}

\subsection{Diversity}

Figure~\ref{fig:diversity} shows an increase in \textit{distinct-$n$} when using artist ID embedding regardless of the value of $n$. Despite the better specificity of track IDs, the artist ID embedding model produces more lexically diverse outputs. This indicates that the high \texttt{\textless{}UNK\textgreater} token ratio in track ID (Table~\ref{tab:unk}) lowers not only semantic relevance but also diversity performance.



\begin{table*}[]
\centering
\resizebox{\linewidth}{!}{%
\begin{tabular}{l|l|l}
\toprule
PID & 143802 & 75870 \\ \midrule
\begin{tabular}[c]{@{}l@{}} {$F_t$} \end{tabular} & 355.39 & 0.75 \\ \midrule
\begin{tabular}[c]{@{}l@{}}Ground Truth\end{tabular} & \begin{tabular}[c]{@{}l@{}}카페를 가득채우는 감성노래\\ (An atmosphere of touching music permeates a cafe.)\end{tabular} & \begin{tabular}[c]{@{}l@{}}나른한 오후에 어울리는 잔잔한 연주\\ (Calming melodies in the afternoon when feeling drowsy)\end{tabular} \\ \midrule
\begin{tabular}[c]{@{}l@{}}Track ID Embedding \end{tabular} & \begin{tabular}[c]{@{}l@{}}따뜻한 커피한잔과 함께 듣는 발라드\\ (Ballad music and a cup of coffee in hand))\end{tabular} & \begin{tabular}[c]{@{}l@{}}영화 속 ost 모음\\ (Movie soundtracks compilation)\end{tabular} \\ \midrule
\multirow{2}{*}{\begin{tabular}[c]{@{}l@{}}Artist ID Embedding\end{tabular}} & \multirow{2}{*}{\begin{tabular}[c]{@{}l@{}}센치한 밤에 듣는 감성 노래\\ (A sentimental night with touching music)\end{tabular}} & \multirow{2}{*}{\begin{tabular}[c]{@{}l@{}}잔잔한 피아노 선율로 듣는 클래식\\ (Piano melodies in classical music that are soothing)\end{tabular}} \\
 &  &  \\ \bottomrule
\end{tabular}
}
\caption{\label{tab:inference}Inference examples from the highest 25\% $F_t$ group (\textbf{Left}) and the lowest 25\% $F_t$ group (\textbf{Right}).}
\end{table*}

\begin{table*}[]
\resizebox{\linewidth}{!}{%
\begin{tabular}{l|l|l}
\toprule
PID & 117646 & 116152 \\ \midrule
\begin{tabular}[c]{@{}l@{}} {$F_a$}\end{tabular} & 3161.39 & 540.40 \\ \midrule
\begin{tabular}[c]{@{}l@{}}Ground Truth\end{tabular} & \begin{tabular}[c]{@{}l@{}}달빛아래 들려오는 감미로운 보이스\\ (A sweet voice in the moonlight)\end{tabular} & \begin{tabular}[c]{@{}l@{}}한국 힙합 흐름 따라가기\\ (Catching up with the flow of the Korean hip-hop scene)\end{tabular} \\ \midrule
\begin{tabular}[c]{@{}l@{}}Track ID Embedding\end{tabular} & \begin{tabular}[c]{@{}l@{}}따뜻한 커피한잔과 함께 듣는 발라드\\ (Ballad music and a cup of coffee in hand)\end{tabular} & \begin{tabular}[c]{@{}l@{}}영화 속 bgm rock 영화 속 bgm 모음\\ (BGM rock music in the movie, Compilation of movie BGMs)\end{tabular} \\ \midrule
\multirow{2}{*}{\begin{tabular}[c]{@{}l@{}}Artist ID Embedding\end{tabular}} & \multirow{2}{*}{\begin{tabular}[c]{@{}l@{}}조용한 밤에 듣는 감성 노래\\ (Touching songs you might wanna listen to on a calm night)\end{tabular}} & \multirow{2}{*}{\begin{tabular}[c]{@{}l@{}}한국 힙합의 시작에서 정착까지\\ (Korean hip-hop: From its inception until its maturation)\end{tabular}} \\
 &  &  \\ \bottomrule
\end{tabular}
}
\caption{\label{tab:inference2}Inference examples from the highest 25\% $F_a$ group (\textbf{Left}) and the lowest 25\% $F_a$ group (\textbf{Right}).}
\end{table*}

\subsection{Robustness}
We assessed the model's robustness by comparing its performance when dealing with infrequently appearing data versus frequently appearing data.
We defined the frequency of appearances from two perspectives: average track frequency and average artist frequency. Based on these points of view, we equally divided the entire playlist into four groups and compared the model's performance on the top 25\% and bottom 25\%. In this paper, we denote them as $F_t$ and $F_a$, respectively.
The average track frequency $F_t$ and the average artist frequency $F_a$ are calculated in the following way. When a playlist consists of tracks ($t_1, ..., t_m$) and artists ($a_1, ..., a_n$), the average track frequency {$F_t$} is obtained by taking the mean of track frequency ($f_1, ..., f_m$). Similarly, we can calculate the average artist frequency score {$F_a$} by averaging the frequency of each artist in the train sets ($\hat{f}_1, ..., \hat{f}_n$) (The numbers of $m$ and $n$ could be different because there could be multiple artists on a single track).
\begin{equation} 
\label{track_popularity}
F_t = \frac{1}{m}\Sigma_{x=1}^{m} {f_x}
\end{equation}

\begin{equation} 
\label{artist_popularity}
F_a = \frac{1}{n}\Sigma_{x=1}^{n} {\hat{f}_x}
\end{equation}

As shown in figure \ref{fig:popularity_distribution}, there were a much greater number of playlists with low $F_t$ than those with low $F_a$. 

The artist ID embedding model shows remarkably better performance on the lowest $F_t$ score group (those from 0\% to 25\%) than the track ID embedding in terms of semantic relevance, as shown in figure \ref{fig:bert_score_f_t}. The improvement is made without impairing the performance on the highest $F_t$ score group (those from 75\% to 100\%). In fact, the performance on the highest $F_t$ group improves when using artist ID embedding, as in the case of the Million Playlist Dataset. As a consequence of this, the performance gap between the subgroups is mitigated.
We also observed that the artist ID embedding improved the performance on the lowest $F_a$ score group (those from 0\% to 25\%) and alleviated the performance gap, as shown in Table \ref{fig:bert_score_f_a}. 

\subsection{Qualitative Results}
We provide four inference examples on playlists from the Melon Playlist Dataset, along with their PIDs (unique identification numbers assigned to each playlist) in Table \ref{tab:inference} and \ref{tab:inference2}, with English translation. Each playlist is from the group that has the highest 25\% $F_t$, the lowest 25\% $F_t$, the highest 25\% $F_a$, and the lowest 25\% $F_a$, respectively. Regardless of the inputs, the model successfully generated a phrase-like title describing the mood (e.g., \textit{touching}) or location (e.g., \textit{cafe}) of the playlists from the highest 25\% $F_t$ and highest 25\% $F_a$ group. The model trained with artist ID embedding produced informative titles that represent the mood (\textit{soothing}), originality (\textit{Korean}) and genre (\textit{hip-hop}) of the playlists from the lowest 25\% $F_t$ and lowest 25\% $F_a$ group. However, when trained with track IDs, the model could not generate a title with the correct information for the playlist from the lowest 25\%  $F_t$ and lowest 25\% $F_a$ groups (PID: 75870). Whereas the ground truth defines each playlist as \textit{calming melodies} and \textit{Korean hip-hop}, the model predicts them as \textit{movie soundtracks} and \textit{rock music in the movie}, respectively.


\section{Conclusions}
In this paper, we proposed a model for music playlist title generation using a series of artist IDs as input. Throughout our assessment, we showed that the artist ID embedding can improve the performance in terms of word overlap, semantic similarity, and diversity. We also proved that the artist ID embedding improves the model's robustness because rarely or never-before-seen data are less likely to be handled by the model. Our findings were supported by two datasets in different languages. However, our research has two limitations. First, in the real world, a single artist might cover multiple genres and styles over time. A fixed artist ID embedding does not reflect such variety. Second, the artist ID embedding model would still need to deal with unseen data (new artists). One approach to solving this problem is to use content-based audio embedding for encoder input and observe the model's performance. For future work, we plan to experiment with different inputs such as audio, genre labels, or a mixture of multiple inputs.

\bibliography{aaai23.bib}



\end{document}